\documentclass[aps,prb,twocolumn,showpacs,groupedaddress, amsmath]{revtex4}

\usepackage{graphicx}
\usepackage{amssymb}

\begin{document}


\title{Study on successive superconducting transitions in Ta$_{2}$S$_{2}$C from electrical resistivity and nonlinear AC magnetic susceptibility}


\author{Masatsugu Suzuki}
\email[]{suzuki@binghamton.edu}
\affiliation{Department of Physics, State University of New York at Binghamton, Binghamton, New York 13902-6000}

\author{Itsuko S. Suzuki}
\email[]{itsuko@binghamton.edu}
\affiliation{Department of Physics, State University of New York at Binghamton, Binghamton, New York 13902-6000}

\author{Takashi Noji}
\email[]{noji@teion.apph.tohoku.ac.jp}
\affiliation{Department of Applied Physics, Graduate School of Engineering, Tohoku University, Sendai, 980-8579 JAPAN}

\author{Yoji Koike}
\email[]{koike@teion.apph.tohoku.ac.jp}
\affiliation{Department of Applied Physics, Graduate School of Engineering, Tohoku University, Sendai, 980-8579 JAPAN}

\author{J\"{u}rgen Walter}
\email[]{Juerg_Walter@t-online.de}
\affiliation{Department of Materials Science and Processing, Graduate School of Engineering, Osaka University, 2-1, Yamada-oka, Suita, 565-0879, JAPAN}


\date{\today}

\begin{abstract}
Ta$_{2}$S$_{2}$C compound undergoes superconducting transitions at $T_{cl} = 3.60 \pm 0.02$ K and $T_{cu} = 9.0 \pm 0.2$ K. The nature of successive superconducting transitions has been studied from electrical resistivity, linear and nonlinear AC magnetic susceptibilities. The resistivity $\rho$ at $H$ = 0 shows a local maximum near $T_{cu}$, a kink-like behavior around $T_{cl}$, and reduces to zero at below $T_{0}$ = 2.1 K. The $\ln T$ dependence of $\rho$ is observed at $H$ = 50 kOe at low temperatures, which is due to two-dimensional weak-localization effect. Below $T_{cu}$ a two-dimensional superconducting phase occurs in each TaC layer. The linear and nonlinear susceptibilities $\chi_{1}^{\prime\prime}$, $\chi_{3}^{\prime}$, $\chi_{5}^{\prime}$, and $\chi_{7}^{\prime}$ as well as the difference $\delta\chi$ ($= \chi_{FC} - \chi_{ZFC}$) between the FC and ZFC susceptibilities, start to appear below 6.0 K, the onset temperature of irreversibility. A drastic growth of the in-plane superconducting coherence length below 6.0 K gives rise to a three-dimensional superconducting phase below $T_{cl}$, through interplanar Josephson couplings between adjacent TaC layers. The oscillatory behavior of $\chi_{3}^{\prime\prime}$, $\chi_{5}^{\prime\prime}$, and $\chi_{7}^{\prime\prime} $ below $T_{cl}$ is related to the nonlinear behavior arising from the thermally activated flux flow. 
\end{abstract}


\maketitle



\section{\label{intro}INTRODUCTION}
Ta$_{2}$S$_{2}$C has a unique layered structure, where a sandwiched structure of C-Ta-S-vdw-S-Ta-C is periodically stacked along the $c$ axis. A van der Waals (vdw) gap is between adjacent S layers.\cite{ref01,ref02} The structure of Ta$_{2}$S$_{2}$C can be viewed as a structural sum of TaC and TaS$_{2}$ layers. The structural part corresponding to the TaS$_{2}$ layer is identical to the atom disposition of either 1T-TaS$_{2}$ in the case of 3R-Ta$_{2}$S$_{2}$C or a hypothetical 2H$_{b}$-TaS$_{2}$ (MoS$_{2}$-type) in the case of 1T- Ta$_{2}$S$_{2}$C (see the schematic diagram of the structure of Ta$_{2}$S$_{2}$C elsewhere\cite{ref02,ref03}). The bulk TaC shows a superconductivity around 9 K,\cite{ref19,ref20} while the bulk 1T-TaS$_{2}$ shows an Anderson localization effect at low temperatures.\cite{ref12,ref08,ref13} Our interest lies in the electrical and magnetic properties of Ta$_{2}$S$_{2}$C which will be discussed in association with the quasi-2D character of the system. 

In our previous paper\cite{ref01} we have undertaken an extensive study on the DC and AC magnetic susceptibility of 3R-Ta$_{2}$S$_{2}$C. We have shown that this compound undergoes successive superconducting phase transitions at $T_{cl}$ ($= 3.60 \pm 0.02$ K) and $T_{cu}$ ($= 9.0 \pm 0.2$ K). Between $T_{cu}$ and $T_{cl}$ the two-dimensional (2D) superconducting phase mainly occurs in each TaC layer. Below $T_{cl}$, there occurs a three-dimensional (3D) superconducting phase through interplanar Josephson couplings between adjacent TaC layers. We have also shown that the localized magnetic moments of conduction electrons associated with the Anderson localization appear probably in the 1T-TaS$_{2}$ layers. 

In the present paper we examine (i) the nature of the successive superconducting phase transitions at $T_{cu}$ and $T_{cl}$, (ii) the 2D weak localization effect in a strong magnetic field, and (iii) the $H$-$T$ phase diagram. To this end, we measure the electrical resistivity $\rho$ of Ta$_{2}$S$_{2}$C in the absence and the presence of an external DC magnetic field $H$. We also measure the linear and nonlinear AC magnetic susceptibilities ( $\chi_{2n+1}^{\prime}$ and $\chi_{2n+1}^{\prime\prime}$ with $n$ = 0, 1, 2, 3) in the absence of $H$. When the magnetic behavior of a system is completely reversible and the electric field ($E$) vs current density ($J$) relationship is linear, there will be no higher-order harmonics associated with the AC magnetic susceptibility. However, when the $E$-$J$ relationship is nonlinear, there will be a non-linear $M$-$H$ response, leading to the generation of higher-order harmonics. Such a nonlinearity of the system arises from the presence of magnetic flux pinning. 

We show that the resistivity strongly depends on the temperature $T$ and $H$. The resistivity $\rho$ at $H$ = 0 shows a local maximum near $T_{cu}$ and a kink-like behavior around $T_{cl}$. It reduces to zero below $T_{0}$ = 2.1 K. A 2D superconducting phase occurs in each TaC layer below $T_{cu}$. We show that the irreversible effect of magnetization occurs below 6.0 K from the linear and nonlinear susceptibilities $\chi_{1}^{\prime\prime}$, $\chi_{3}^{\prime}$, $\chi_{5}^{\prime}$, and $\chi_{7}^{\prime}$ as well as the difference $\delta\chi$ ($= \chi_{FC}-\chi_{ZFC}$), where $\chi_{FC}$ and $\chi_{ZFC}$ are field-cooled (FC) and zero-field cooled (ZFC) DC magnetic susceptibilities. The drastic growth of the in-plane superconducting coherence length below 6.0 K gives rise to a 3D superconducting phase below $T_{cl}$, through interplanar Josephson couplings between adjacent TaC layers. The $H$-$T$ phase diagram is also discussed. We show that the oscillatory behavior of $\chi_{3}^{\prime\prime}$, $\chi_{5}^{\prime\prime}$, and $\chi_{7}^{\prime\prime}$ below $T_{cl}$ is related to the nonlinear behavior arising from the thermally activated flux flow. We show that the $\ln T$ dependence of $\rho$ at $H$ = 50 kOe at low temperatures is due to the 2D weak localization effect. Our results from the electrical resistivity and nonlinear AC magnetic susceptibility will be compared with those from DC magnetic susceptibility and linear AC magnetic susceptibility which have been already reported in our previous paper.\cite{ref01}

\section{\label{exp}EXPERIMENTAL PROCEDURE}
Powdered samples of Ta$_{2}$S$_{2}$C were prepared by Pablo Wally.\cite{ref04,ref05,ref06} The detail of the synthesis and structure is described in previous paper.\cite{ref01} X-ray powder diffraction pattern shows that Ta$_{2}$S$_{2}$C sample consists of a 3R phase as a majority phase and a 1T phase as a minority phase. The electrical resistivity was measured using PPMS (Quantum Design) in the temperature range ($0.5 \leq T \leq 298$ K) and the magnetic field ($0 \leq H \leq 50$ kOe). The sample has a form of rectangular-prism pellet ($3.5 \times 0.8 \times 6.5$ mm$^{3}$) prepared from polycrystalline powdered Ta$_{2}$S$_{2}$C by pressing it. The four-probe method was used for the measurement. The voltage probes and current probes were attached to the surface of the sample by using silver paste. The DC magnetic susceptibility was measured using SQUID magnetometer (Quantum Design, MPMS XL-5). The detail of the cooling protocol process for the measurement of the ZFC and FC susceptibilities is described in previous paper.\cite{ref01} 

The nonlinear AC magnetic susceptibility of Ta$_{2}$S$_{2}$C was measured by using SQUID magnetometer (Quantum Design MPMS XL-5) with AC susceptibility option. When the AC magnetic field [$= h \cos (\omega t)$] at angular frequency $\omega$ ($= 2\pi f$) is applied, the AC magnetization $M(\omega ,t)$ induced in the system is expressed by a sum of Fourier components with odd angular frequency $(2n+1)\omega$ ($n$ = 0, 1, 2, 3,..... ),\cite{ref07}
\begin{equation}
M(\omega ,t)=\sum\limits_{n=0}\{ \Theta_{2n+1}^{\prime} \cos [(2n+1)\omega t] +\Theta_{2n+1}^{\prime\prime} \sin [(2n+1)\omega t] \},
\label{eq01}
\end{equation}
where 
\begin{equation}
\Theta_{1}^{\prime}=\chi_{1}^{\prime}h+\frac{3}{4} \chi_{3}^{\prime}h^{3}+\frac{5}{8}\chi_{5}^{\prime}h^{5}+\frac{35}{64}\chi_{7}^{\prime}h^{7}+\frac{63}{128} \chi_{9}^{\prime}h^{9}+\cdots ,
\label{eq02}
\end{equation}
and
\begin{equation}
\Theta_{3}^{\prime}=\frac{1}{4}\chi_{3}^{\prime}h^{3}+\frac{5}{16}\chi_{5}^{\prime}h^{5}+\frac{21}{64}\chi_{7}^{\prime}h^{7}+\frac{21}{64}\chi_{9}^{\prime} h^{9}+\cdots ,
\label{eq03}
\end{equation}
and similarly for the imaginary part $\Theta_{2n+1}^{\prime\prime}$:
\begin{equation}
\Theta_{1}^{\prime\prime}=\chi_{1}^{\prime\prime}h+\frac{3}{4}\chi_{3}^{\prime\prime}h^{3}+\frac{5}{8}\chi_{5}^{\prime\prime}h^{5}+\frac{35}{64}\chi_{7}^{\prime\prime}h^{7}+\frac{63}{128}\chi_{9}^{\prime\prime}h^{9}+\cdots ,
\label{eq04}
\end{equation}
and
\begin{eqnarray}
&\Theta_{2n+1}^{\prime\prime}&(h,\chi_{1}^{\prime\prime},\chi_{3}^{\prime\prime},\chi_{5}^{\prime\prime},\cdots)\nonumber \\
&=&(-1)^{n}\Theta_{2n+1}^{\prime}(h,\chi_{1}^{\prime} \rightarrow \chi_{1}^{\prime\prime},\chi_{3}^{\prime} \rightarrow \chi _{3}^{\prime\prime},\chi_{5}^{\prime} \rightarrow \chi_{5}^{\prime\prime},\cdots) .\nonumber \\
\label{eq05}
\end{eqnarray}

The linear and nonlinear AC magnetic susceptibilities were measured in the following way. Before setting up a sample at 298 K, a remnant magnetic field in a superconducting magnet was reduced to less than 3 mOe using an ultra low field capability option. The system was cooled from 298 to 2.0 K at zero-magnetic field (ZFC protocol). In our AC SQUID system, only the real part $\Theta_{1}^{\prime}$ and the imaginary part $\Theta_{1}^{\prime\prime}$ of the AC magnetization were simultaneously measured as a function of the amplitude of AC magnetic field $h$ ($0.05 \leq h \leq 4$ Oe) at $H$ = 0 at each $T$ ($2.0 \leq T \leq 12.0$ K). The frequency of the AC magnetic field was kept at $f$ = 0.1 Hz. After each $h$-scan measurement at the fixed $T$, the temperature was increased by $\Delta T$ = 0.1 K and the $h$-scan measurements were repeated. The linear and nonlinear susceptibilities $\chi_{2n+1}^{\prime}$ and $\chi_{2n+1}^{\prime\prime}$ with $n$ = 0, 1, 2, and 3 are determined from the least squares fits of the data to Eq.(\ref{eq02}) for $\Theta_{1}^{\prime}$ and to Eq.(\ref{eq04}) for $\Theta_{1}^{\prime\prime}$. Note that this method is suitable for the nonlinear susceptibility for spin glass systems at low frequencies ($f = 0.01 - 0.1$ Hz), where the nonlinear susceptibility may diverge at a spin freezing temperature in the absence of a DC magnetic field. 

\section{\label{result}RESULT}
\subsection{\label{resultA}Electrical resistivity}

\begin{figure}
\includegraphics[width=7.0cm]{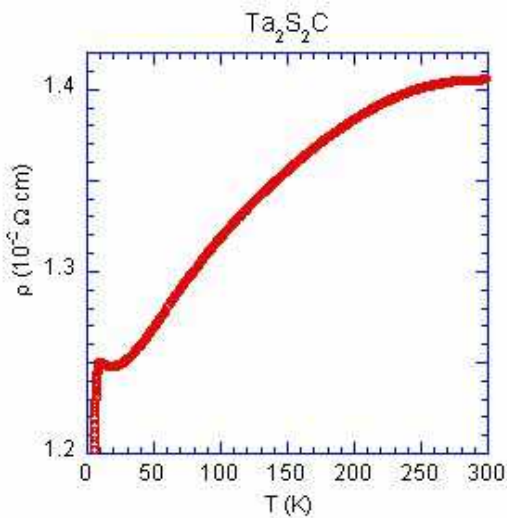}
\caption{\label{fig01}(Color online) $T$ dependence of the resistivity $\rho$ at $H$ = 0 for Ta$_{2}$S$_{2}$C. }
\end{figure}

\begin{figure}
\includegraphics[width=7.0cm]{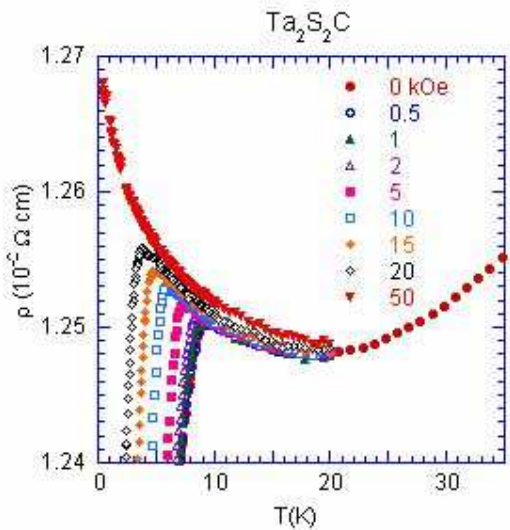}
\caption{\label{fig02}(Color online) $T$ dependence of $\rho$ at fixed $H$ (= 0, 0.5, 1, 2, 5, 10, 15, 20, and 50 kOe). }
\end{figure}

\begin{figure}
\includegraphics[width=7.0cm]{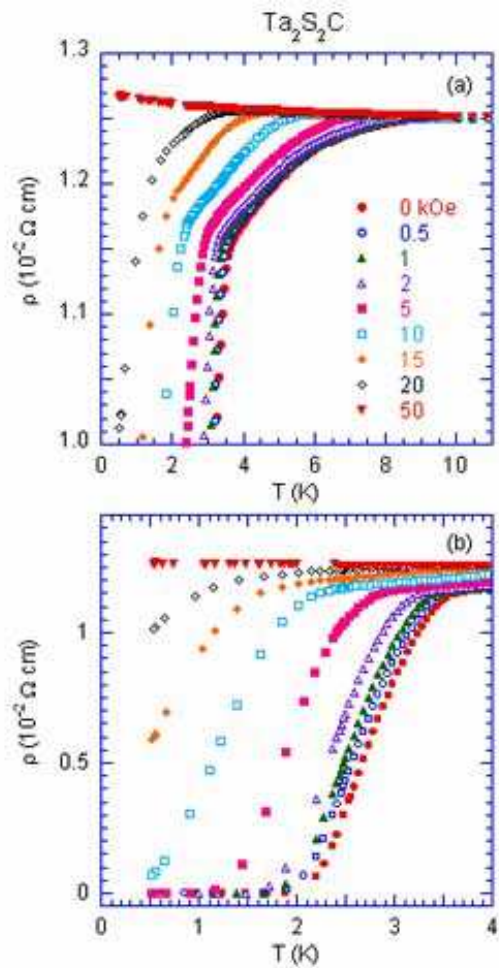}
\caption{\label{fig03}(Color online) (a) and (b) $T$ dependence of $\rho$ at low temperatures for various $H$. }
\end{figure}

We have measured the electrical resistivity as a function of $T$ and $H$. Figure \ref{fig01} shows the $T$ dependence of the electrical resistivity $\rho$ at $H$ = 0. In Figs.~\ref{fig02} and \ref{fig03}, for clarity we also show the $T$ dependence of $\rho$ at various magnetic field $H$ for $0 \leq H \leq 50$ kOe, depending on the ranges of $T$ and the values of $\rho$. The resistivity $\rho$ at $H$ = 0 decreases with decreasing $T$ from 300 K, showing a metallic behavior at high temperatures: As shown in Fig.~\ref{fig02}, $\rho$ at $H$ = 0 shows a local minimum at $T_{min}$ (= 19 K), and starts to increase with decreasing $T$, showing a bit of evidence for the possible localization effect. The resistivity $\rho$ at $H$ = 0 exhibits a local maximum at $T_{max}$ ($\approx 9.6$ K) and again starts to decreases with decreasing $T$. This local-maximum temperature $T_{max}$ is close to $T_{cu}$. As shown in Figs.~\ref{fig03}(a) and (b), $\rho$ at $H$ = 0 exhibits a kink-like behavior at $T_{kink}$ ($\approx 3.6$ K). This kink temperature corresponds to $T_{cl}$. The resistivity $\rho$ drastically decreases with decreasing $T$ below $T_{kink}$ and reduces to zero below $T_{0}$ = 2.1 K, suggesting that the superconductivity occurs at low $T$ over the whole system. For convenience, we also define the characteristic temperature $T_{1/2}$ at which $\rho$ is equal to a half of the normal resistivity at $T_{max}$. As shown in Figs.~\ref{fig02} and \ref{fig03}, the curve of $\rho$ vs $T$ is strongly dependent on $H$. The characteristic temperatures $T_{max}$, $T_{kink}$, $T_{1/2}$ and $T_{0}$ decrease with increasing $H$. The $H$-$T$ diagram of these temperatures will be described in Sec.~\ref{resultD} and is compared with that of $T_{cu}(H)$ and $T_{cl}(H)$.\cite{ref01} In Sec.~\ref{dis0} we will show that $\rho$ at $H$ = 50 kOe is proportional to $\ln T$ at low temperatures. This behavior will be discussed in association with a 2D weak localization effect.

\subsection{\label{resultB}Linear and nonlinear AC magnetic susceptibilities}

\begin{figure}
\includegraphics[width=7.0cm]{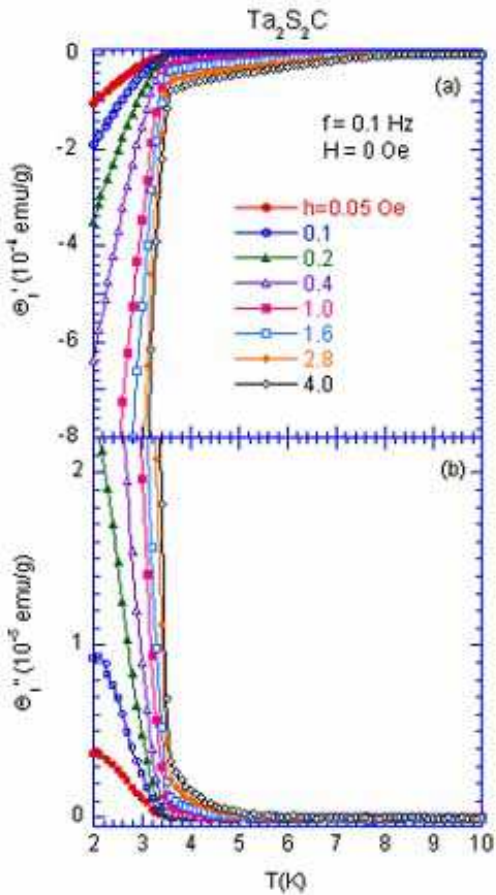}
\caption{\label{fig05}(Color online) $\Theta_{1}^{\prime}$ vs $T$ and (b) $\Theta _{1}^{\prime\prime}$ vs $T$ at various amplitude of AC magnetic field, $h = 0.05 - 4$ Oe. $f$ = 0.1 Hz. The solid lines are guides to the eyes.}
\end{figure}

\begin{figure*}
\includegraphics[width=12.0cm]{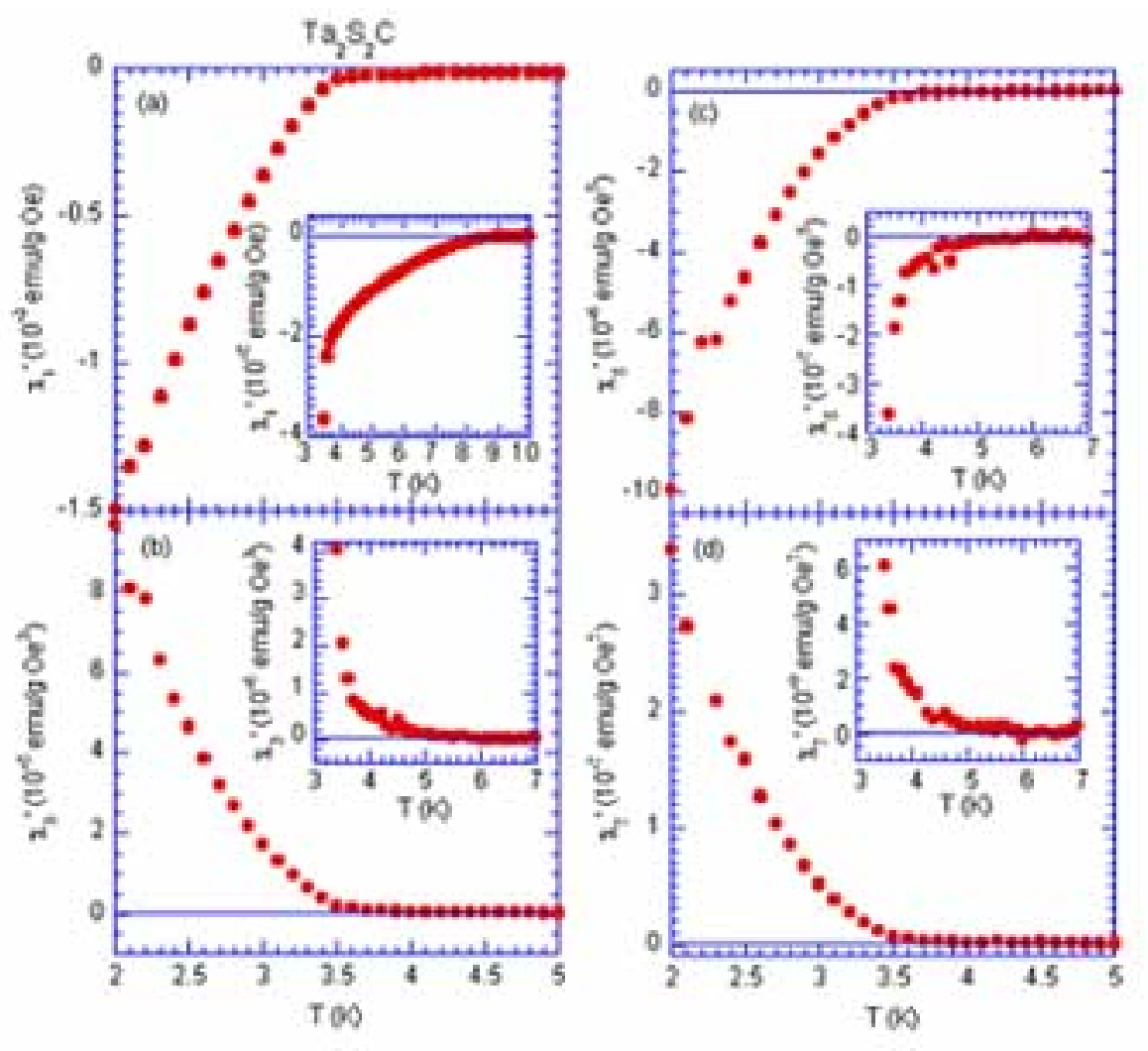}
\caption{\label{fig06}(Color online) $T$ dependence of linear and nonlinear susceptibilities (a) $\chi_{1}^{\prime}$, (b) $\chi_{3}^{\prime}$, (c) $\chi_{5}^{\prime}$, and (d) $\chi_{7}^{\prime}$. $f$ = 0.1 Hz. These are determined from the least-squares fit of the data $\Theta_{1}^{\prime}$ vs $h$ for $0.05 \leq h \leq 4.0$ Oe at each $T$ to Eq.(\ref{eq02}). In the insets, we show the detail of the $T$ dependence of $\chi_{1}^{\prime}$, $\chi_{3}^{\prime}$, $\chi_{5}^{\prime}$, and $\chi_{7}^{\prime}$. }
\end{figure*}

\begin{figure*}
\includegraphics[width=12.0cm]{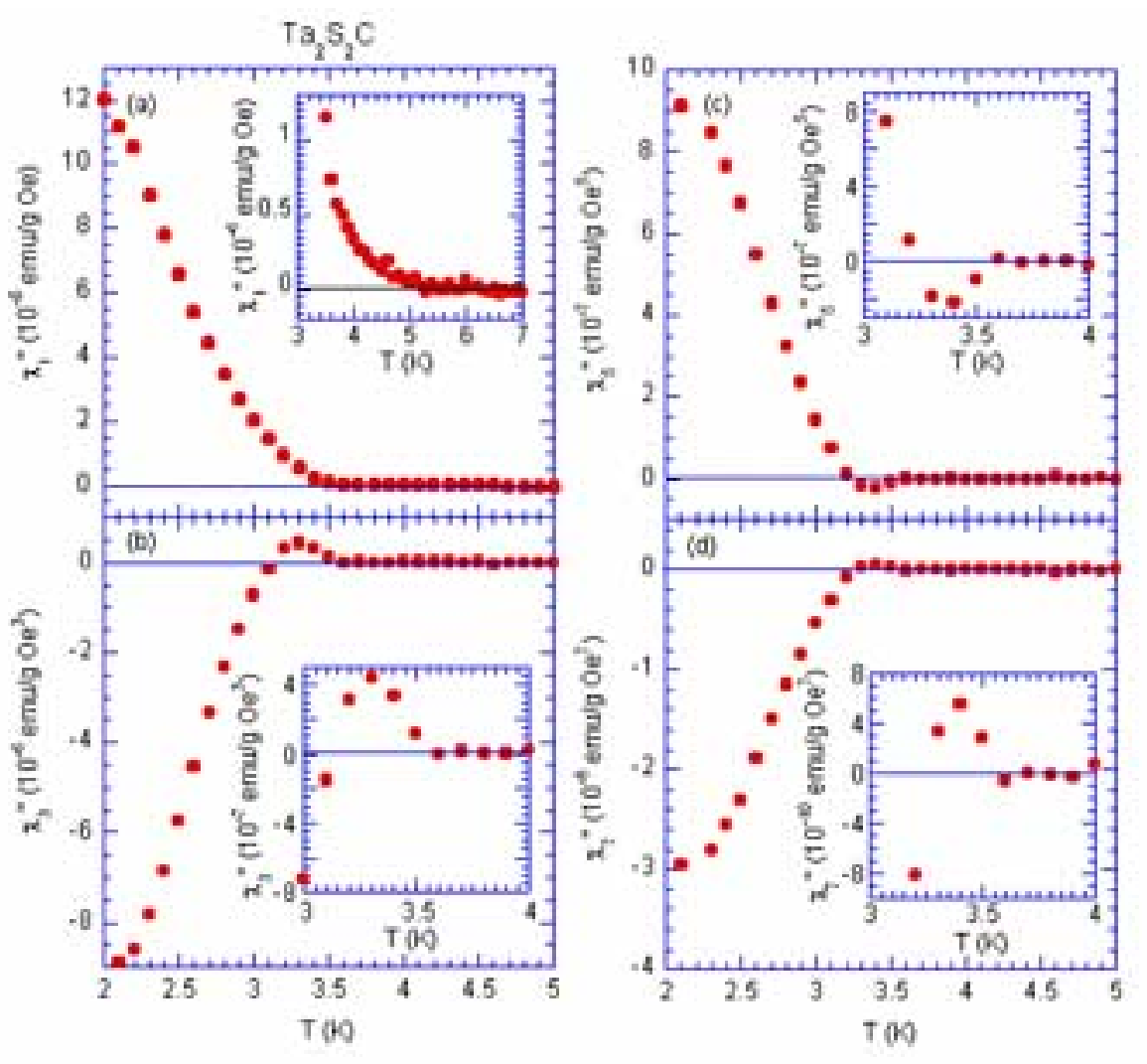}
\caption{\label{fig07}(Color online) $T$ dependence of linear and nonlinear susceptibilities (a) $\chi_{1}^{\prime\prime}$, (b) $\chi_{3}^{\prime\prime}$, (c) $\chi_{5}^{\prime\prime}$, and (d) $\chi_{7}^{\prime\prime}$. $f$ = 0.1 Hz. These are determined from the least-squares fit of the data $\Theta_{1}^{\prime}$ vs $h$ for $0.05 \leq h \leq 4.0$ Oe at each $T$ to Eq.(\ref{eq04}). In the insets, we show the detail of the $T$ dependence of $\chi_{1}^{\prime\prime}$, $\chi_{3}^{\prime\prime}$, $\chi_{5}^{\prime\prime}$, and $\chi_{7}^{\prime\prime}$ around $T_{cl}$. }
\end{figure*}

We have determined the $T$ dependence of the linear and nonlinear AC magnetic susceptibilities of Ta$_{2}$S$_{2}$C. Figure \ref{fig05} shows the $T$ dependence of $\Theta_{1}^{\prime}$ and $\Theta_{1}^{\prime\prime}$ at various AC magnetic fields ($0.05\leq h\leq 4$ Oe) in the absence of a DC magnetic field, where $f$ = 0.1 Hz. It is experimentally confirmed that both $\Theta_{1}^{\prime}$ and $\Theta_{1}^{\prime\prime}$ are well described by a series expansion of only odd power of $h$ [see Eqs.(\ref{eq02}) and (\ref{eq04})]. The $T$ dependence of the real parts $\chi_{1}^{\prime}$, $\chi_{3}^{\prime}$, $\chi_{5}^{\prime}$, and $\chi_{7}^{\prime}$ and the imaginary parts $\chi_{1}^{\prime\prime}$, $\chi_{3}^{\prime\prime}$, $\chi_{5}^{\prime\prime}$, and $\chi_{7}^{\prime\prime}$ are shown in Figs.~\ref{fig06} and \ref{fig07}, respectively. The results are summarized as follows. (i) $\chi_{1}^{\prime}$ is negative below $T_{cu}$. It shows a kink-like behavior at $T_{cl}$ and reduces to zero at $T_{cu}$. (ii) Both $\chi_{3}^{\prime}$ and $\chi_{7}^{\prime}$ are positive at low temperatures. They show a kink-like behavior at $T_{cl}$ and reduce to zero around 6.0 K. Note that the positive sign of $\chi_{3}^{\prime}$ is observed below $T_{c}$ in melt grown YBa$_{2}$Cu$_{3}$O$_{7-\delta}$ by Polichctti et al.\cite{ref14} ($f$ = 10.7 Hz. $H$ = 0, $h$ = 4 Oe). (iii) $\chi_{5}^{\prime}$ is negative at low temperatures. It shows a kink-like behavior at $T_{cl}$ and reduces to zero around 6.0 K. (iv) $\chi_{1}^{\prime\prime}$ is positive at low temperatures. It shows a kink-like behavior at $T_{cl}$ and reduces to zero around 6.0 K. Note that the $T$ dependence of $\chi_{1}^{\prime\prime}$ is similar to that of $\chi_{3}^{\prime}$. (v) $\chi_{3}^{\prime\prime}$ is negative at low temperatures. It reduces to zero at 3.1 K, showing a positive peak at 3.3 K, and reduces to zero around $T_{cl}$. Similar oscillatory behavior in $\chi_{3}^{\prime\prime}$ vs $T$ below the transition temperature $T_{c}$ has been reported by Ishida et al.\cite{ref15} in a high $T_{c}$-superconductor (Sr$_{0.7}$Ca$_{0.3}$)$_{0.95}$CuO$_{2-x}$. (vi) $\chi_{5}^{\prime\prime}$ is positive at low temperatures. It reduces to zero at 3.2 K, showing a negative local minimum at 3.4 K, and reduces to zero around 3.6 K. (vii) $\chi_{7}^{\prime\prime}$ is negative at low temperatures. It crosses the zero-line around 3.2 K, showing a positive peak at 3.5 K, and reduces to zero around $T_{cl}$. The physical meaning of the $T$ dependence of $\chi_{2n+1}^{\prime}$ and $\chi_{2n+1}^{\prime\prime}$ ($n$ = 0, 1, 2, 3) will be discussed in Secs.~\ref{disA} and \ref{disB} in association with the onset of irreversibility in magnetization below 6.0 K and the presence of thermally activated flux flow below $T_{cl}$. 

\subsection{\label{resultC}$\chi_{ZFC}$ and $\chi_{FC}$}

\begin{figure}
\includegraphics[width=7.0cm]{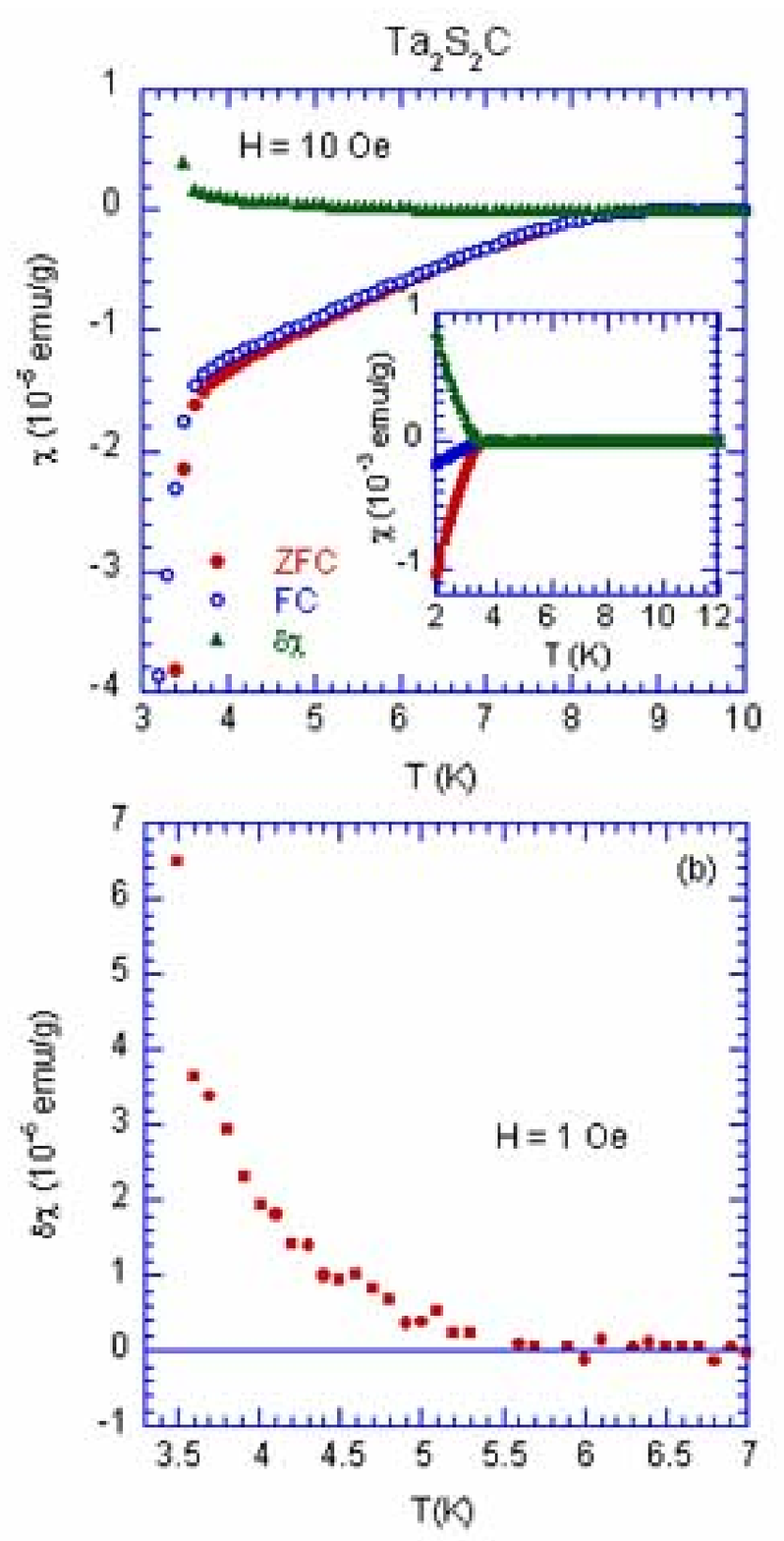}
\caption{\label{fig08}Color online) (a) $T$ dependence of the zero-field cooled susceptibility $\chi_{ZFC}$, the field-cooled susceptibility $\chi_{FC}$, and the difference $\delta\chi$ (= $\chi_{FC}-\chi_{ZFC}$) for $H$ = 10 Oe. (b) Detail of the $T$ dependence of $\delta\chi$ near 6.0 K. $H$ = 1 Oe. }
\end{figure}

Our results on the ZFC susceptibility $\chi_{ZFC}$ and the FC susceptibility $\chi_{FC}$ have been reported in detail in our previous paper.\cite{ref01} These results are consistent with the present results derived from the measurement of $\rho$ vs $T$ (Sec.~\ref{resultA}). Here we summarize our results using two figures which are not published in the previous paper.\cite{ref01} The method of the measurements is the same as that in the previous paper.\cite{ref01} Figure \ref{fig08}(a) shows the $T$ dependence of $\chi_{ZFC}$, $\chi_{FC}$, and the difference $\delta\chi$ at $H$ = 10 Oe, where $\delta\chi=\chi_{FC}-\chi_{ZFC}$. Figure \ref{fig08}(b) shows the $T$ dependence of $\delta\chi$ at $H$ = 1 Oe. The difference starts to appear below 6.0 K between $T_{cl}$ and $T_{cu}$. This temperature is defined as an onset-temperature of irreversibility in magnetization at $H$ = 1 Oe. 

\begin{figure}
\includegraphics[width=7.0cm]{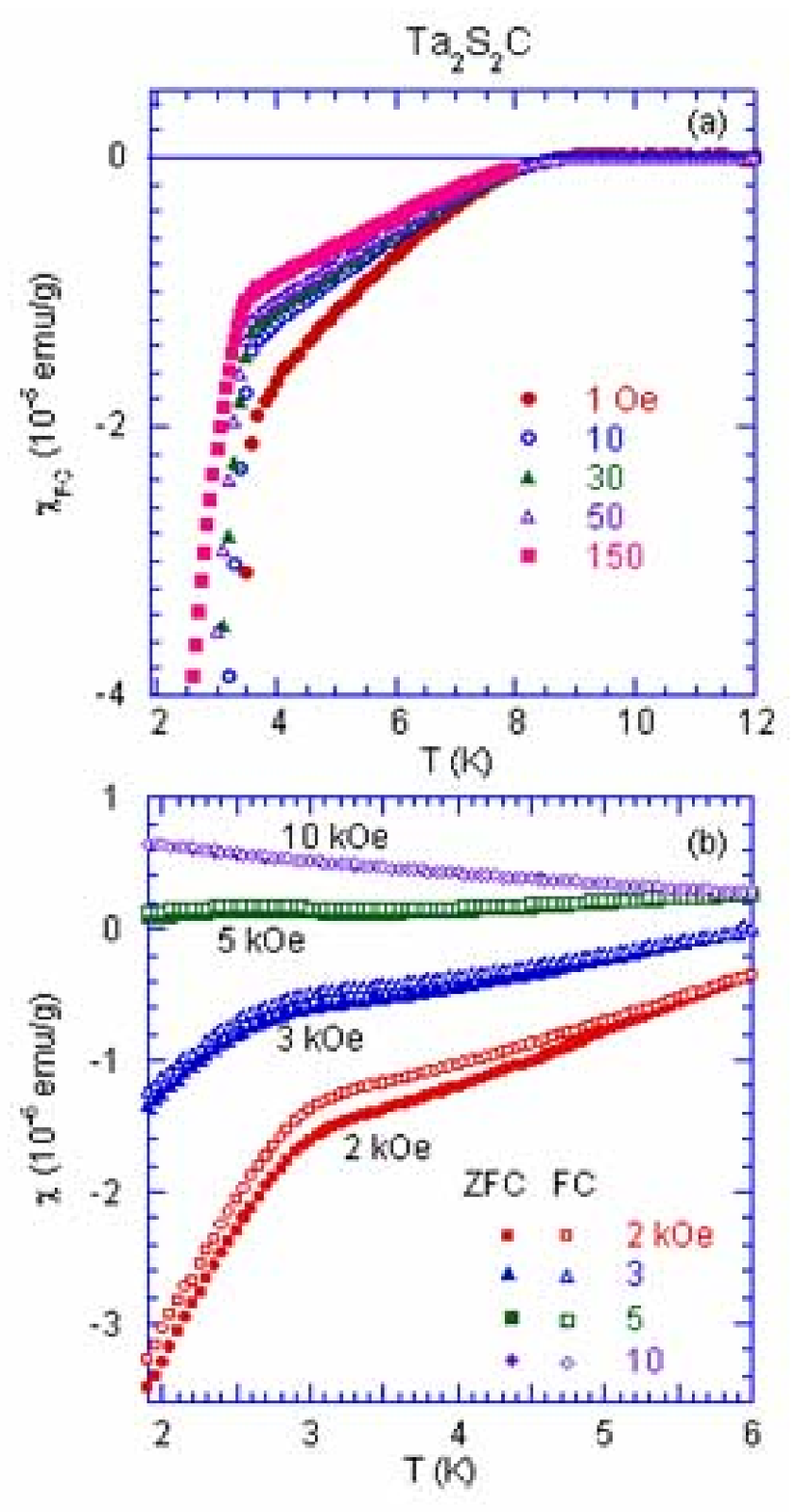}
\caption{\label{fig09}(Color online) (a) $T$ dependence of $\chi_{FC}$ at $H$ = 1, 10, 30, 50, and 150 Oe. (b) $T$ dependence of $\chi_{FC}$ and $\chi_{FC}$ at $H$ = 2, 3, 5, and 10 kOe. }
\end{figure}

Figure \ref{fig09}(a) shows the $T$ dependence of $\chi_{FC}$ for $1 \leq H \leq 150$ Oe. Figure  \ref{fig09}(b) shows the $T$ dependence of $\chi_{ZFC}$ and $\chi_{FC}$ for $H\geq 2$ kOe. In Fig.~\ref{fig08}(a) and Fig.~\ref{fig09}(a), the susceptibility $\chi_{ZFC}$ ($\chi_{FC}$) exhibits a kink at $T_{cl}(H)$, where d$\chi_{ZFC}$/d$T$ (d$\chi_{FC}$/d$T$) undergoes a discontinuous jump. The deviation of $\chi_{ZFC}$ from $\chi_{FC}$ is clearly seen below $T_{cl}(H)$, indicating that the extra magnetic flux is trapped during the FC process (see more detail of $\delta\chi$ vs $T$ at various $H$ in our previous paper\cite{ref01}). As shown in Fig.~\ref{fig09}(b), there is a drastic decrease in the diamagnetic contribution in $\chi_{ZFC}$ with increasing $T$ below $T_{cl}(H)$. Nevertheless, a diamagnetic contribution in $\chi_{ZFC}$ still remains above $T_{cl}(H)$, increases with further increasing $T$, and becomes zero at a upper critical temperature $T_{cu}(H)$. The sign of $\chi_{ZFC}$ changes from negative to positive around 9 K with increasing $T$.\cite{ref01} At $H$ = 5 kOe, $\chi_{ZFC}$ is positive at least between 1.9 and 6 K, showing a broad peak at 2.65 K. At $H$ = 10 kOe, $\chi_{ZFC}$ decreases with increasing $T$, showing a Curie-like behavior.\cite{ref01} We note that the susceptibility at low $T$ for 1T-TaS$_{2}$ layer shows a Curie-like behavior due to the localized magnetic moments of conduction electrons related to the Anderson localization effect.\cite{ref10} 

\subsection{\label{resultD}$H$-$T$ phase diagram}

\begin{figure}
\includegraphics[width=7.0cm]{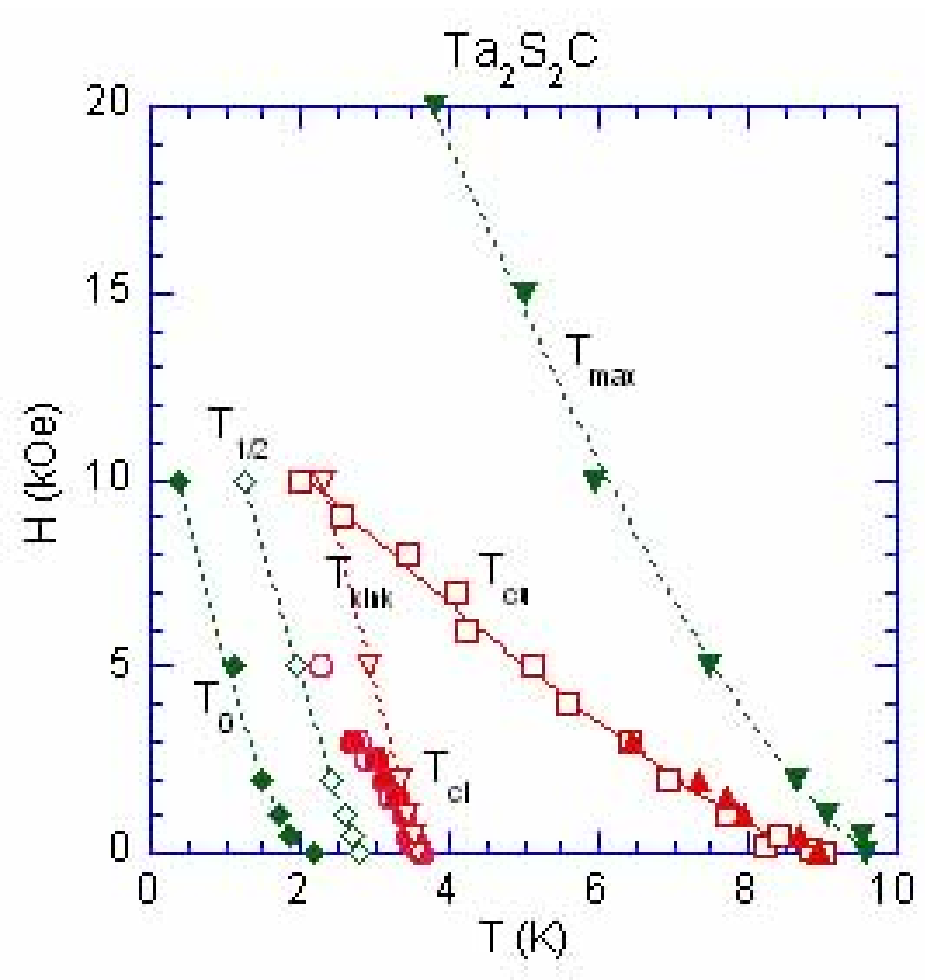}
\caption{\label{fig10}(Color online) $H$-$T$ phase diagram. The characteristic temperatures ($T_{0}$, $T_{1/2}$, $T_{kink}$, $T_{max}$) are defined from the resistivity measurement; $T_{0}$ where $\rho$ reduces to zero, $T_{1/2}$ where $\rho$ is equal to one half of the normal resistivity, $T_{kink}$ where $\rho$ shows a kink-like behavior, and $T_{max}$ where $\rho$ exhibits a local maximum. The critical temperatures ($T_{cu}$ and $T_{cl}$) are defined from the DC and AC magnetic susceptibility,\cite{ref01} $T_{cl}$ is determined from the measurements of $\Theta_{1}^{\prime}/h$ ($f$ = 1 Hz and $h$ = 0.5 Oe) vs $T$ and $\Theta_{1}^{\prime\prime}/h$ ($f$ = 1 Hz and $h$ = 0.5 Oe) vs $T$. $T_{cu}$ is determined from the measurements of $\Theta_{1}^{\prime}/h$ ($f$ = 1 Hz and $h$ = 0.5 Oe) vs $T$ and $\chi_{ZFC}$ vs $T$. The solid line is least-squares fitting curve for the data of $T_{cu}(H)$ to Eq.(\ref{eq08}). The dotted lines are guide to the eyes.}
\end{figure}

Figure \ref{fig10} shows the $H$-$T$ phase diagram. Here $T_{cl}(H)$ is determined from the data of $\Theta_{1}^{\prime}/h$ vs $T$ and $\Theta_{1}^{\prime\prime}/h$ vs $T$ at $f$ = 1 Hz and $h$ = 0.5 Oe, and $T_{cu}(H)$ is determined from the data of $\chi_{ZFC}$ vs $T$ (see the previous paper\cite{ref01} in detail). The negative sign of $\chi_{ZFC}$ and $\chi_{FC}$, and $\Theta_{1}^{\prime} /h$ below $T_{cu}(H)$ indicates that the system is at least partially in a superconducting state. The dispersion $\Theta_{1}^{\prime}/h$ shows a kink-like behavior at $T_{cl}(H)$. Four characteristic temperatures are determined from the resistivity measurement: (i) $T_{0}$ where $\rho$ reduces to zero (the real Meissner state), (ii) $T_{1/2}$ where $\rho$ is equal to one half of the normal resistivity (this is conventionally used as the definition of the superconducting transition temperature in the resistivity measurement of high $T_{c}$ superconductors\cite{ref16,ref17,ref18}), (iii) $T_{kink}$ where $\rho$ shows a kink-like behavior (a crossover behavior occurs from the 2D superconducting phase to the 3D superconducting phase), and (iv) $T_{max}$ where $\rho$ exhibits a local maximum (a part of TaC clusters become superconductivity). Note that this definition of $T_{max}$ is also used as the onset temperature of superconductivity for the electron-doped high-$T_{c}$ superconductors (granular superconductors).\cite{ref16,ref17,ref18} As shown in Fig.~\ref{fig10}, the line $T_{kink}(H)$ almost coincides with the line $T_{cl}(H)$. The line $T_{max}(H)$ is near the line $T_{cu}(H)$ at low $H$ but the difference between them becomes larger as $H$ increases. This may be due to the superconducting fluctuations sensitively observed by the resistivity measurements. The lines $T_{0}(H)$ and $T_{1/2}(H)$ are a little lower than the line $T_{cl}(H)$ at the same $H$ by $1.2 - 1.5$ K. This difference may be reduced when the structural coupling between fine particles in the pelletized sample becomes strong during the pressing process. It seems that the lines $T_{cu}(H)$ and $T_{kink}(H)$ merge into a point located at $H_{m}$ = 10 kOe and $T_{m}$ = 2 K in the $H$-$T$ plane. In the previous paper\cite{ref01} we assume that the line $T_{cu}(H)$ corresponds to the upper critical fields $H_{c2}^{(u)}(T)$. The line $H_{c2}^{(u)}(T)$ can be well described by 
\begin{equation}
H_{c2}^{(u)}(T)=H_{c2}^{(u)}(T=0)(1-\frac{T}{T_{cu}})^{\alpha} ,
\label{eq08}
\end{equation}
with $\alpha = 1.23 \pm 0.07$, $T_{cu} = 9.0 \pm 0.2$ K, and $H_{c2}^{(u)}(T=0)=14.5 \pm 0.5$ kOe.\cite{ref01} 

We note that the superconducting phase between $T_{cu}$ and $T_{cl}$ in Ta$_{2}$S$_{2}$C is similar to that observed in bulk TaC,\cite{ref19,ref20} TaC nanoparticles based on carbon nanotube,\cite{ref21} TaC encapsulated into the inner core of carbon nanotubes,\cite{ref22} and Ta-metal graphite based on natural graphite.\cite{ref23} 

(i) Bulk TaC with a rock salt structure shows type-II superconductivity where $T_{c}$ = 10.2 K, the lower critical field $H_{c1}=220$ Oe, and the upper critical field $H_{c2}=4.6$ kOe.\cite{ref20} (ii) In TaC nanoparticles (carbon nanotube) synthesized at $1000 - 1100^\circ$C, the size of TaC nanoparticles is on the order of $270 - 340$ $\AA$. This compound shows superconductivity with a critical temperature $T_{c}$ (= 10.5 K) and critical field $H_{c2}$ (= 16.3 kOe).\cite{ref21} (iii) The ZFC and FC susceptibility measurements on TaC crystal encapsulated into the inner core of carbon nanotubes,\cite{ref22} reveal that this system exhibits a superconductivity below 10 K. The resistivity $\rho$ shows a metallic behavior for $T>$60 K. The $T$ dependence of $\rho$ for $13<T<60$ K is well described by a 3D variable range hopping model; $\ln (\rho /T^{1/2})$ is proportional to $1/T^{1/4}$. (iv) In Ta-metal graphite, Ta metal layer is sandwiched between adjacent graphene sheets, forming a periodic stacking sequence along the $c$ axis.\cite{ref23} This compound shows the superconductivity below $T_{c}$ = 10 K at $H$ = 0. The critical field $H_{c2}$ increases with increasing $T$ for $H_{c2}\leq 0.3$ kOe, showing a reentrant phase. By this phase one mean that at fixed $T$ close to 10 K the system changes from the normal ($N$) phase to the superconducting (S) phase and from the S phase to the $N$ phase with increasing $H$. The critical field $H_{c2}$ increases with decreasing $T$ for $H_{c2}>0.5$ kOe, and reaches 20 kOe at 6 K. 
Thus it can be concluded from these results that the superconductivity below $T_{cu}$ in Ta$_{2}$S$_{2}$C occurs in each TaC layer.

\section{\label{dis}DISCUSSION}
\subsection{\label{dis0}2D weak localization effect}
It is interesting to compare our results of $\rho$ vs $T$ at $H$ = 50 kOe with that at $H$ = 0 and 50 kOe for bulk 1T-TaS$_{2}$.\cite{ref08} The overall behavior of $\rho$ vs $T$ at $H$ = 50 kOe is qualitatively similar to that of bulk 1T-TaS$_{2}$ at least below the charge density wave (CDW) commensurate transition temperature $T_{d}$ ($\approx 200$ K).\cite{ref08} The resistivity $\rho$ at $H$ = 0 for bulk 1T-TaS$_{2}$ shows a local minimum around 50 K and increases with further decreasing $T$. The value of $\rho$, which is between 30 and 40 m$\Omega$cm, is relatively larger than that ($\approx$ 12.5 m$\Omega$cm) for $\rho$ at $H$ = 50 Oe for Ta$_{2}$S$_{2}$C (see Fig.~\ref{fig02}). Note that our result of $\rho$ vs $T$ is rather different from that of Ta$_{2}$S$_{2}$C reported by Ziebarath et al.\cite{ref09} using a pressed-and sintered sample. They have observed that $\rho$ increases with increasing $T$ for $4.2 \leq T \leq 300$ K ($\rho \approx$ 0.8 m$\Omega$cm at 4.2 K and 2.2 m$\Omega$cm at 300 K), showing a metallic behavior. No discontinuous change in $\rho$ has been seen below 300 K. 

\begin{figure}
\includegraphics[width=7.0cm]{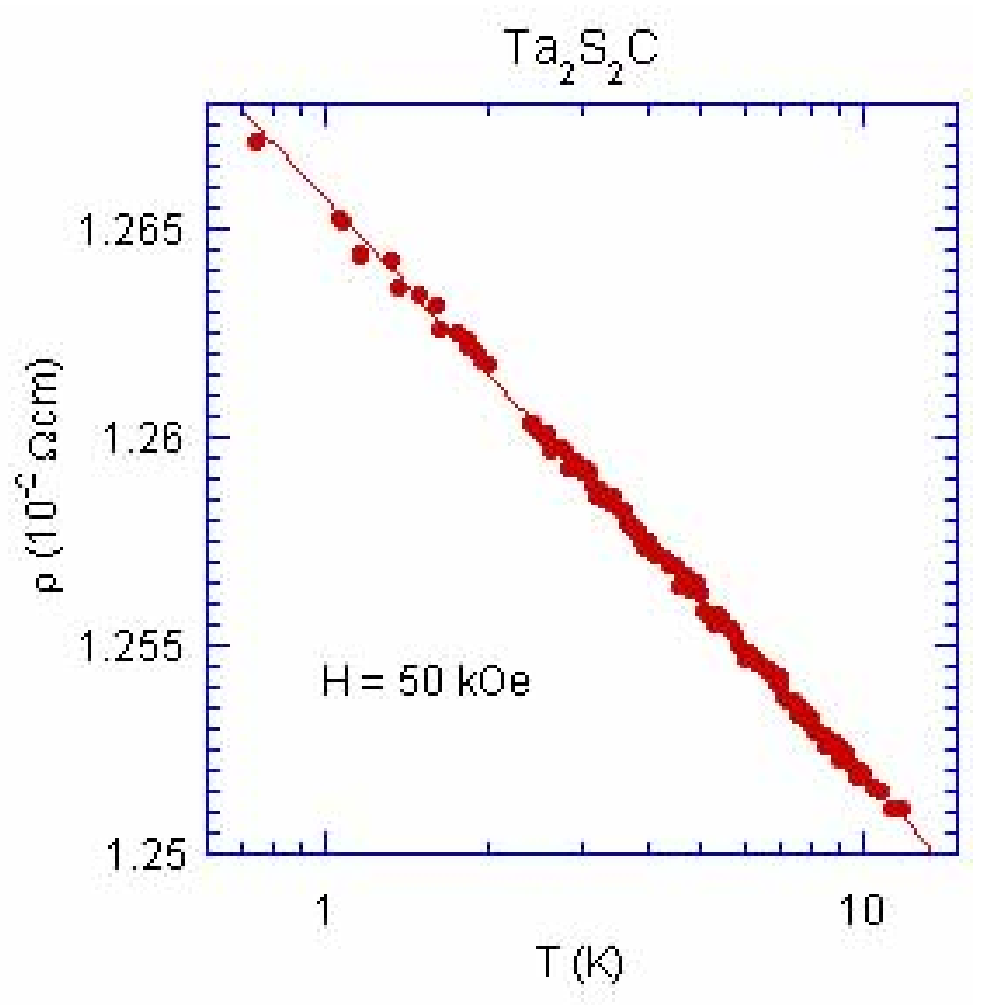}
\caption{\label{fig04}(Color online) $T$ dependence of resistivity at $H$ = 50 kOe for $0.7 \leq T \leq 12$ K. The solid line denotes the least-squares fitting curve to Eq.(\ref{eq06}). }
\end{figure}

It seems that the local minimum temperature $T_{min}(H)$ for $\rho$ vs $T$ is independent of $H$ for $0 \leq H \leq 50$ kOe (see Fig.~\ref{fig02}). The curve of $\rho (H,T)$ vs $T$ for $0 \leq H \leq 20$ kOe deviates from that at $H$ = 50 kOe at low temperatures below $T_{max}(H)$. No local maximum in $\rho$ vs $T$ is observed at $H$ = 50 kOe: $\rho$ continues to increase with decreasing $T$ at least down to $T$ = 0.5 K. Figure \ref{fig04} shows the plot of $\rho$ at $H$ = 50 kOe as a function of a logarithmic scale of $T$ for $0.7 \leq T \leq 12$ K. We find that $\rho$ is well described by 
\begin{equation}
\rho=\rho_{0}-\rho_{1}\ln T ,
\label{eq06}
\end{equation}
with $\rho_{0}$ = 12.66 m$\Omega$cm and $\rho_{1} = 5.99 \times 10^{-2}$ m$\Omega$cm. 
Note that the value of $\rho_{1}$ is rather different from that predicted from the 2D weak localization, partly because our system is a powdered sample. No evidence for the negative magnetoresistance is observed in Ta$_{2}$S$_{2}$C at temperatures above 0.5 K. However, there is some possibility that the negative magnetoresistance may be observed at much lower temperatures. In fact, for 1T-Ta$_{2}$, the negative resistance is observed only below 100 mK, while the positive magnetoresistance is observed above $T$ = 200 mK.\cite{ref13}
Nevertheless, the above result suggests that the mutual interaction effect between electrons in a 2D weakly localized state\cite{ref10} occurs in Ta$_{2}$S$_{2}$C at $H$ = 50 kOe. Similar behavior has been reported in electron-doped high-$T_{c}$ superconductor Nd$_{2-x}$Ce$_{x}$CuO$_{y}$ with $x$ = 0.15.\cite{ref11} The superconducting state is destroyed above around $H$ = 100 kOe even at 1.3 K. The normal resistivity which are obtained by extrapolating the high-field value to zero-field value, increases with decreasing $T$, showing a $\ln T$ dependence.  

Here we show that the 2D weak localization observed in Ta$_{2}$S$_{2}$C is closely related to the Anderson localization effect in bulk 1T-TaS$_{2}$. For bulk 1T-TaS$_{2}$, the resistivity $\rho$ below 2 K diverges according to the relation
\begin{equation}
\rho=\rho_{2}+\rho_{3}\exp [(T_{a}/T)^{1/\zeta}] ,
\label{eq07}
\end{equation}
where $\zeta$ ($=d+1$) is an exponent, $d$ is the dimension of the system, $\rho_{2}$ and $\rho_{3}$ are constant resistivities, and $T_{a}$ is a characteristic temperature. The exponent $\zeta$ is equal to 3, which has been first reported by DiSalvo and Graebner.\cite{ref12} The value of $\zeta$ = 3 is consistent with the 2D variable range hopping conduction. Furukawa et al.\cite{ref13} have measure the electrical resistivity of bulk 1T-TaS$_{2}$ prepared in several ways as a function of $T$ and $H$. The $T$ dependence of $\rho$ at $H$ = 0 for 40 mK $\leq T \leq 4$ K is well described by Eq.(\ref{eq07}) with $\zeta =3.0$, $T_{a} = 1.08 \pm 0.19$ K, $\rho_{3} = 9.72 \pm 1.84$ m$\Omega$cm, and $\rho_{2} = 2.66 \pm 3.99$ m$\Omega$cm, indicating that the 2D variable range hopping occurs at $H=0$. In contrast, the $T$ dependence of $\rho$ at $H$ = 50 kOe for 40 mK $\leq T \leq 4$ K is well described by Eq.(\ref{eq06}) with $\rho_{1} = 11.58 \pm 0.23$ m$\Omega$cm and $\rho_{0} = 37.49 \pm 0.40$ m$\Omega$cm. This result indicates that the 2D weak localization effect occurs in bulk 1T-TaS$_{2}$ at $H=50$ kOe. 
When $H$ is applied, the energy difference between the Fermi level and the mobility edge of electrons with spins parallel to the filed direction becomes small, leading to a marked increase of the hopping probability of the electrons from one site to another site and to a marked decrease in the resistivity.\cite{Fukuyama1979} 

No 2D variable range hopping conduction ($\zeta = 3$) occurs at $H$ = 0 in Ta$_{2}$S$_{2}$C since the system shows superconductivity.
The $\ln T$ dependence of $\rho$ observed at $H$ = 50 kOe in Ta$_{2}$S$_{2}$C is due to the 2D weak localization effect in the 1T-TaS$_{2}$ layers. However, the normal state of the TaC layers is also expected to contribute to the 2D electronic conduction. Therefore, it is concluded that the $\ln T$ dependence of $\rho$ is due to the 2D weak localization effect of both 1T-TaS$_{2}$ layers and TaC layers.

\subsection{\label{disA}Onset of irreversibility}
In our previous paper,\cite{ref01} we have shown that the difference $\delta\chi$ ($= \chi_{FC}-\chi_{ZFC}$) shows a kink at $T_{cl}$ = 3.60 K. However, it does not reduce to zero at $T=T_{cl}$, but exhibit a broad tail above $T_{cl}$. In Fig.~\ref{fig08}(b) we show that $\delta\chi$ at $H=1$ Oe reduces to zero around 6.0 K. It is assumed that the onset temperature of irreversibility $T_{irr}$ is equal to 6.0 K. Above this temperature, $\chi_{ZFC}$ is exactly equal to $\chi_{FC}$. The linear susceptibility $\chi_{1}^{\prime}$ shows a kink at $T_{cl}$. It reduces to zero at $T_{cu}$. This susceptibility is a measure of the AC flux penetration (AC magnetic-field screening), rather than the irreversibility. We find that $\chi_{1}^{\prime\prime}$, $\chi_{3}^{\prime}$, $(-\chi_{5}^{\prime})$, and $\chi_{7}^{\prime}$ are positive at low temperatures. They decrease with increasing $T$ and reduce to zero around 6.0 K. This implies that the onset-temperature for $\chi_{1}^{\prime\prime}$, $\chi_{3}^{\prime}$, $(-\chi_{5}^{\prime})$, and $\chi_{7}^{\prime}$ is the same as that for $\delta\chi$. Our results are consistent with the prediction from the Bean critical state model\cite{ref24} that $\chi_{1}^{\prime\prime}$ = 0 and $\chi_{3}^{\prime}$ = 0 in the temperature region where each penetrating quantum flux moves reversibly with the change of small AC field ($H$ = 0). When the magnetic behavior of the system is reversible and $E$-$J$ relationship is linear, there will be no higher-order harmonics. However, when $E$-$J$ relationship is nonlinear, there will be a nonlinear, $M$-$H$ response and the generation of higher order harmonics. When the magnetic behavior is irreversible, it follows that the $E$-$J$ relationship must be nonlinear. The in-plane superconducting coherence length starts to grow below 6.0 K and drastically increases with decreasing $T$ as $T$ is approached $T_{cl}$. Such a drastic growth of the in-plane coherence length $\xi$ gives rise to a 3D superconducting phase below $T_{cl}$ through interplanar Josephson couplings ($J^{\prime}_{inter}$) between adjacent TaC layers. The effective interplanar interaction estimated as $J^{\prime}_{inter}(\xi /a_{0})^{2}$ is on the order of $k_{B}T$ at $T=T_{cl}$, where $a_{0}$ is the in-plane lattice constant of the TaC layer.

\subsection{\label{disB}Thermally activated flux flow below $T_{cl}$}
As is described in Sec.~\ref{resultB}, the nonlinear susceptibilities $\chi_{3}^{\prime\prime}$, $(-\chi_{5}^{\prime\prime})$, and $\chi_{7}^{\prime\prime}$ have a local positive maximum at 3.4 K and a negative local minimum at 3.4 K just below $T_{cl}$. Such an oscillatory behavior of $\chi_{3}^{\prime\prime}$, $(-\chi_{5}^{\prime\prime})$, and $\chi_{7}^{\prime\prime}$ below $T_{cl}$ is related to the nonlinear behavior arising from the thermally activated flux flow. Below $T_{cl}$, it is assume that the resistivity $\rho (B,J)$ which depends on the current density $J$ and the local magnetic field $B$, is formed of the parallel connection between the flux creep ($\rho_{cr}$) and flux flow ($\rho_{ff}$) resistivities (Gioacchino et al.\cite{ref25}). When $JU_{p}(T)/J_{c}(T)k_{B}T\ll 1$, the ratio $E/J$ is independent of $J$, so that the resistivity $\rho_{cr}$ describes a regime with ohmic behavior, where $J_{c}(T)$ is the critical current density at $T$ and $U_{p}(T)$ is the activation energy for flux jumps. When $J$ is close to $J_{c}(T)$ and $U_{p}(T)/k_{B}T\gg 1$, the resistivity $\rho_{cr}(J)$ is dependent on $J$. This describes a non-ohmic regime where a flux creep occurs. In a linear $E$-$J$ characteristic, higher-order harmonics should be absent. However, the magnetic field dependence of the two resistivities $\rho_{cr}$ and $\rho_{ff}$ can induce even in the presence of only AC magnetic field, due to the nonlinear behavior of the system. In this case, higher harmonic components are present in the AC magnetic susceptibility. Gioacchino et al.\cite{ref25} have made numerical calculations on the $T$ dependence of $\chi_{3}^{\prime}$ and $\chi_{3}^{\prime\prime}$ near $T_{c}$ for high-$T_{c}$ superconductors, where $U_{p}(T)$ and $J_{c}(T)$ are different for different pinning models. The imaginary part at low frequencies $\chi_{3}^{\prime\prime}$ shows an oscillatory behavior; it displays a positive value on approaching $T_{c}$ from the low $T$ side, and a positive value at lower temperatures. Thus the oscillatory behavior of $\chi_{3}^{\prime\prime}$, $(-\chi_{5}^{\prime\prime})$, and $\chi_{7}^{\prime\prime}$ below $T_{cl}$ in Ta$_{2}$S$_{2}$C arises from the nonlinear $E$-$J$ relationship through the thermally activated flux flow. 

\subsection{\label{disC}$T_{cl}$ and $T_{cu}$}
We discuss the relation between $T_{cu}$ and $T_{cl}$ in terms of a model proposed by Jardim et al.\cite{ref26} for the granular superconductor Sm$_{2-x}$Ce$_{x}$CuO$_{4-y}$. For simplicity, we assume that our system is a randomly coupled 3D system, where the superconducting grains are all the same size, distributed on a diluted simple cubic network of lattice parameter $a$. A fraction $p$ of the sites of the network is occupied by grains, which are Josephson coupled. With $p > p_{c}$ ($p_{c}$ is a percolation threshold), the Josephson junctions form an infinite connected cluster. The Josephson coupling energy at $T$ is given by\cite{ref26}
\begin{equation}
E_{J}(T)=\frac{23.153}{2\pi}k_{B}T_{cu}(1-T/T_{cu})\frac{R_{c}}{R_{J}}  , 
\label{eq09}
\end{equation}
where $R_{c}=\hbar /e^{2} = 4.11386$ k$\Omega$ and $R_{J}$ is the resistance in the normal state of the junction.  Here we assume that the superconducting phase transition at $T_{cl}$ is that of the 3D XY system; $k_{B} T_{cl}=2.201E_{J}(T_{cl})$. Then the ratio $T_{cl}/T_{cu}$ is derived as\cite{ref26}
\begin{equation}
\frac{T_{cl}}{T_{cu}}=\frac{1}{1+0.123(R_{J}/R_{c})} .
\label{eq10}
\end{equation}
The substitution of $T_{cu}$ = 9.0 K and $T_{cl}$ = 3.60 K in Ta$_{2}$S$_{2}$C into Eq.(\ref{eq10}) leads to $R_{J}$ = 49.94 k$\Omega$. For the percolation model described above, the resistivity in the normal phase is expressed by\cite{ref26} 
\begin{equation}
\rho =R_{J}a(\frac{p-p_{c}}{1-p_{c}})^{-t} ,
\label{eq11}
\end{equation}
where $p_{c}$ is equal to 0.31 for the cubic lattice and $t$ ($\approx 1.9$) is the 3D conductivity exponent. When $\rho$ = 12.5 m$\Omega$cm at 20 K in Ta$_{2}$S$_{2}$C, the lattice constant $a$ is estimated as $a$ = 21.70 $\AA$ for $p$ = 0.95, 18.59 $\AA$ at $p$ = 0.9, 13.06 $\AA$ at $p$ = 0.80, 8.47 $\AA$ at $p$ = 0.70, and 4.82 at $p$ = 0.60. The lattice constant $a$ at $p$ = 0.702 is equal to the separation distance between adjacent TaC layers (8.537 $\AA$) in Ta$_{2}$S$_{2}$C. This value of $a$ is much shorter than we expect. However it is on the same order as that derived by Jadim et al.\cite{ref26} for Sm$_{2-x}$Ce$_{x}$CuO$_{4-y}$. 

We note that a high-$T_{c}$ superconductor ceramics YB$_{2}$Cu$_{4}$O$_{8}$ also undergoes successive superconducting transitions at $T_{cu}$ (= 80 K) and $T_{cl}$ (= 37 K).\cite{ref27,ref28,ref29,Hagiwara2005} The $T$ dependence of $\rho$ at $H=0$, and $\chi_{ZFC}$ and $\chi_{FC}$ at $H=0.1$ Oe is very similar to that observed in Ta$_{2}$S$_{2}$C. The ratio $T_{cu}/T_{cl}$ (= 2.16) is a little smaller than the ratio (= 2.5) for Ta$_{2}$S$_{2}$C. A negative nonlinear magnetic susceptibility and a negative nonlinear electrical resistivity are observed at $T_{cl}$ in the absence of $H$. These behaviors can be explained in terms of a 3D XY chiral glass model,\cite{ref30} where the frustration effect arises from the random distribution of $\pi$ junctions with the negative Josephson coupling. In spite of the similarity between our result and the results on YB$_{2}$Cu$_{4}$O$_{8}$, it seems that our result is not directly related to the 3D XY chiral glass model, because of no anisotropic pairing symmetry of the $d_{x^{2}-y^{2}}$ wave type in our system.  

\subsection{\label{disD}Significance of mesoscopic grains in TaC layers} 
The possible existence of mesoscopic grains in the TaC layers of Ta$_{2}$S$_{2}$C would be essential to the successive transitions having a hierarchical nature. 
Both $\chi_{ZFC}$ and $\chi_{FC}$ are negative below $T_{cu}$. With decreasing $T$, they gradually reduce to zero around $T_{cu}$ in Ta$_{2}$S$_{2}$C. Such a diamagnetic behavior below $T_{cu}$ could be described by a superconductive short range order (SRO) which occurs in each TaC layer. If each conductive layer is perfect or extends infinitely, any frustration could not be expected at all in the ordering process above $T_{cl}$. So we assume here that each conductive layer would be divided into clusters of a finite size. This is mostly the case in a real system because there exist various kinds of defects like impurities, dislocations, twin boundaries, and so on. The universality class of the ordering in 2D superconductor belongs to that of a 2D XY ferromagnet. A Kosterlitz-Thouless type phase transition will appear at $T_{cu}$ (which may be equal to a KT ordering temperature $T_{KT}$).\cite{ref31} Then the correlation length $\xi$ of local superconductive order parameter increases divergingly toward $T_{cu}$ with decreasing $T$ from the high-temperature side. If each 2D layer is not perfect, then the growth of $\xi$ once stops at a certain geometrical length $L_{0}$. Further growth of $\xi$ should then be brought about by some intercluster interaction through the boundary region within each layer. If the intercluster coupling could be a Josephson-type, a Josephson-coupled network among the short range-ordered clusters are realized between $T_{cu}$ and $T_{cl}$. At $T_{cl}$, the ordering characteristic will crossover successfully from the Josephson-coupled 2D network to a Josephson-coupled 3D network. 

\section{CONCLUSION}
The nature of Ta$_{2}$S$_{2}$C ($T_{cl}=3.60 \pm 0.02$ K and $T_{cu} = 9.0 \pm 0.2$ K) has been studied from the measurements of the electrical resistivity and nonlinear AC magnetic susceptibility. The resistivity $\rho$ at $H$ = 0 shows a local maximum near $T_{cu}$ and a kink-like behavior around $T_{cl}$. The intermediate phase between $T_{cu}$ and $T_{cl}$ is a 2D superconducting phase occurring in the TaC layers, while the low temperature phase below $T_{cl}$ is a 3D superconducting phase. The linear and nonlinear susceptibilities $\chi _{1}^{\prime\prime}$, $\chi_{3}^{\prime}$, $\chi_{5}^{\prime}$, and $\chi_{7}^{\prime}$ in the absence of $H$ start to appear below the onset temperature of irreversibility ($T$ = 6.0 K). The drastic growth of the in-plane coherence length below 6.0 K gives rise to the 3D superconducting phase below $T_{cl}$. The oscillatory behavior of $\chi_{3}^{\prime\prime}$, $\chi_{5}^{\prime\prime}$, and $\chi_{7}^{\prime\prime}$ below $T_{cl}$ is related to the nonlinear behavior arising from the thermally activated flux flow. The $\ln T$ dependence of $\rho$ at $H$ = 50 kOe at low temperatures is due to the 2D weak localization effect of the 2D electronic conduction in both 1T-TaS$_{2}$ and TaC layers.

\begin{acknowledgments}
The authors are grateful to Pablo Wally, Technical University of Vienna, Austria (now Littlefuse, Yokohama, Japan) for providing them with the samples. The work at Tohoku University was supported by a Grant-in-Aid for Scientific Research from the Ministry of Education, Culture, Sports, Science and Technology, Japan.
\end{acknowledgments}

\end{document}